\documentclass[%
 reprint,
 amsmath,amssymb,
 aps,
]{revtex4-2}

\usepackage{graphicx}
\usepackage{dcolumn}
\usepackage{bm}
\usepackage{amsmath}
\usepackage{xcolor}
\usepackage{multirow}
\usepackage{hhline}%
\usepackage[section]{placeins}
\usepackage{hyperref}

\begin{document}

\preprint{APS/123-QED}

\title{Surface functionalization modulates collective cell behavior at integer topological defects}


\author{Prasoon Awasthi}
 \email{praw@sdu.dk}
\author{Aniruddh Murali}
\author{Ellen Juel Pørtner}
\author{Adam Cohen Simonsen}
\author{Francesca Serra}
 \email{serra@sdu.dk}
 \affiliation{
 Department of Physics, Chemistry and Pharmacy, University of Southern Denmark.
}

\begin{abstract}

Living cells establish long-range orientational order through collective alignment, giving rise to topological defects whose functional relevance is increasingly recognized in tissue organization and morphogenesis. Engineered topographical patterns have been used to induce such defects in cell monolayers, mimicking natural biological phenomena. In this work, we investigate the effect of cell–surface adhesion on collective cell dynamics at a vortex integer topological defect imposed by a topographical ring pattern. Adhesion strength is controlled via surface functionalization with poly-D-lysine, fibronectin, or covalently bonded fibronectin, and quantified using atomic force microscopy. As surface chemistry is modified, cell morphology changes from irregular to spindle-like, and two distinct collective modes emerge: weakly adhered cells exhibit strong inward motion, while strongly attached cells move tangentially to the ring. Spindle-shaped cells exhibit higher nematic order and promote the emergence of two +1/2 topological defects in the monolayer. We further characterize collective cell dynamics by analyzing correlation lengths and demonstrate the scaling of number density fluctuations in cell systems.

\end{abstract}  

\maketitle


\section{\label{sec:level1}Introduction}
Anisotropic living cells are described as active nematics because of their out-of-equilibrium state and their alignment state, exhibiting quasi-long range orientational order \cite{duclos2017topological,doostmohammadi2018active}. Each cell transduces biochemical energy into motion, and cells altogether generate collective movement \cite{bruckner2025tissue}. Collective cell dynamics is crucial for various physiological processes such as wound healing, morphogenesis, and cancer progression \cite{friedl2009collective}. During these processes, cells self-organize and form patterns, occasionally revealing long-range orientational order \cite{morales2019liquid} and topological defects, which are singularities in the orientation field. Topological defects have been shown to play an important role in the growth of bacterial colonies \cite{doostmohammadi2016, copenhagen2021}, organization of tissues \cite{saw2017}, aggregation/depletion of cells \cite{kawaguchi2017, guillamat2022integer}, and morphogenesis \cite{maroudas2021}. Disclinations of half-integer topological charge ($\pm$1/2) spontaneously emerge in in-vitro systems and have been extensively characterized in terms of dynamics and mechanical properties. Integer defects, however, do not spontaneously form in vitro, but they have been observed as aster, spiral or vortex patterns in many biological systems: vortices and rosettes in epithelial cells \cite{dua1996vortex}, \cite{harding2014roles}, asters in neural rosette \cite{ma2008}, asters and vortices in sclera \cite{gogola2018radial}, spirals in brain tumor \cite{kepes1976cellular} and fibroma \cite{mandal2007fibroma}, and vortices in the plant meristem \cite{hamant2008}.

These defects can be harnessed to influence cell behavior, alignment, and growth, offering a promising strategy for guiding tissue architecture. To incorporate such defects into tissue scaffolds helps mimic the structural and functional organization of native tissues. This approach holds significant potential for clinical applications, including the development of advanced implants that integrate more effectively with host tissue and promote functional recovery. To this end, topological defects with integer charge have been induced in different cells using liquid crystal elastomer (LCE) \cite{turiv2020topology}, topographical pattern \cite{Endresen2021, Kaiyrbekov2023, zhao2024integer, zhao2025integer} or anchoring patterns \cite{maroudas2021}. In every system, the cells (human dermal fibroblasts, 3T6 fibroblasts, neural progenitor cells, and myoblasts) show an inward migration towards the defect core. 

While all these studies focus primarily on the alignment of cells, it is important to note that many of these studies use functionalized surfaces to promote the adhesion of cells on the substrate. This is commonly achieved by treating the surface with proteins or peptides such as fibronectin, collagen, laminin, or others, which determine the adhesion strength. Adhesion molecule complexes that link cells to each other and to the extracellular matrix play a crucial role in the collective motion of cells \cite{eckert2023hexanematic}. In fact, cell motility is reduced as a result of the maturation of the cell-cell and cell-substrate junctions \cite{garcia2015}, impacting not only the cell speed but also the velocity correlation length and the collective motion.

However, cell motility is influenced by additional factors. For example, fibronectin (FN), an ubiquitous extracellular matrix (ECM) protein, has previously been used to strengthen cell-substrate adhesion and to enhance cell motility \cite{tarle2015,ravasio2015, ravasio2015gap}. The colonies of Madin-Darby canine kidneys (MDCK) in patches demonstrate two modes of collective motion (circular and radially outward) at high FN concentration \cite{ravasio2015}. Another study observes faster gap closure in MDCK monolayers on substrates with more FN, where the dominant mode of motility is crawling, primarily driven by lamellipodia protrusion at positively curved edges \cite{ ravasio2015gap}. Surprisingly, rat embryonic fibroblast (REF$_{WT}$) reveals the opposite behavior of migration. The migration speed of a single cell REF$_{WT}$ is reduced with increasing FN concentration, but the directional persistence is enhanced \cite{missirlis2016}. An increase in cell motility is also observed with increased concentration of type I collagen, another ECM component, up to a threshold beyond which motility declines.

Inspired by these studies, in the present research work we show that the collective cell dynamics near defects with integer topological charge is strongly affected by the cell-surface adhesion strength.

\section{\label{sec:level2}Experimental section}

\subsection{Ring pattern substrate fabrication by soft lithography}

For the fabrication of the ring pattern, we follow the procedure previously described in \cite{Endresen2021, Kaiyrbekov2023} to make a master mold of SU-8 photoresist with photolithography. The height, width, and gap between the ridges are 1.5 $\mu$m, 10 $\mu$m, and 60 $\mu$m, respectively. The prepolymer of polydimethylsiloxane (PDMS) from Sylgard 184 (Dow Corning) is prepared in a 10:1 (w/w ratio of monomer to crosslinker). After mixing properly, the prepolymer is desiccated to remove the bubbles. The prepolymer is then poured on top of the patterned SU-8 substrate, and the whole sample is again desiccated to remove any possible bubble trapped between the surfaces. After desiccation, the samples are heated at 60\textdegree C overnight to cure the PDMS, and the cured PDMS is peeled off the SU-8 substrate. 

The patterned PDMS is then used to make a negative mold of UV-curable glue. In this process, a few drops of UV-curable glue Norland optical adhesive 81 (NOA-81) are poured on a cleaned glass petri dish, and the patterned PDMS is placed face-down on top of the UV glue. The sample is desiccated to remove trapped bubbles in the glue, and irradiated with UV at 302 nm (8W, Ultra Violet Products-3UV) for 20 min on each side to harden the glue. After curing the glue, the sample is heated at 60\textdegree C for 30 min. The PDMS is then peeled off, leaving behind the negative replica of the ring pattern on the cured glue. The mold is then used for PDMS replication in lieu of the original SU-8 mold.

To ensure good adhesion between the patterned PDMS slab and the petri dish for the cell experiments, the dish is first coated with a thin PDMS layer. Both the bottom surface of the patterned PDMS and the thin PDMS layer in the petri dish are treated with oxygen plasma (Harrick Plasma Cleaner) for 3 min at 30 W RF power and 250 mtorr pressure. After plasma, the bottom surface of the patterned PDMS is attached to the petri dish and heated at 60\textdegree C for 1 min. After the attachment, the top surface of the PDMS is again treated with oxygen plasma, keeping the same parameters. The samples are then sterilized by submerging them in ethanol for 10 minutes and dried before further chemical treatment.

\subsection{Surface functionalization}
We prepare samples with coatings of poly-D-lysine (PDL, Group I), adsorbed fibronectin (FN, Group II), and covalently bonded FN (xFN, Group III). Group I (PDL) samples are coated with 5 $\mu$g/cm$^2$ PDL (Gibco$^{\text{TM}}$ ThermoFisher) for 30 min at room temperature, then washed with sterile Milli-Q water three times, and left to dry for 90 min. Group II samples are coated with 1 $\mu$g/cm$^2$ fibronectin bovine plasma (Sigma-Aldrich), left overnight at 4\textdegree C, and washed with sterile Milli-Q water three times. Group III samples are prepared following the protocol reported in \cite{kuddannaya,priyadarshani}. Samples are immersed into 10\% of (3-aminopropyl) triethoxysilane (APTES) (Sigma-Aldrich) at 50\textdegree C for 2 hours, dipped into 2.5\% glutaraldehyde (Sigma-Aldrich) at room temperature for 1 hour, and then coated with the fibronectin bovine plasma of 5 $\mu$g/cm$^2$ and left overnight at 4\textdegree C. The samples are washed with sterile Milli-Q water three times after each chemical treatment step, except after removing the fibronectin, when the samples are washed twice with cell culture media at 37\textdegree C. After washing, the samples are completely covered with the media and left in the cell incubator for 4 hours.

\subsection{Cell culture}

NIH 3T3 mouse fibroblast cells (ATCC) are used for the experiments upto passage number 15. The cell line is maintained in the cell incubator at 5\% CO$_2$ and 37\textdegree C in a humidified atmosphere using the Thermo Fisher Scientific Nunclon$^{\text{TM}}$ Delta surface-coated dishes. The cell media consists of 89\% of Dulbecco’s Modified Eagle’s Medium (DMEM)-high glucose formulation (containing 4.5 g/L glucose, L-glutamine, sodium pyruvate, and sodium bicarbonate) (Sigma-Aldrich), 10\% fetal bovine serum (Sigma-Aldrich), and 1\% penicillin-streptomycin solution (Sigma-Aldrich).

After chemically modifying the topographical pattern surface by coating it as described above, the cells are seeded on the patterned surfaces at 200 \#/mm$^2$ density for live cell imaging for cell dynamics and at 20 \#/mm$^2$ density on a plain PDMS substrate for atomic force microscopy (AFM) analysis. We let the cells settle on the pattern in the cell incubator for 40 minutes. After this, the whole petri dish is filled with the media and kept in the incubator overnight.

\subsection{Live and wide-field fluorescence imaging}

The next day after seeding the cells on the pattern and before starting the live imaging, the cell nuclei are stained by adding 1 drop of NucBlue$^{\text{TM}}$ Live Cell ReadyProbes$^{\text{TM}}$ reagent (Hoescht 33342) per ml of cell culture media. The cells are left in the cell incubator for 20 minutes. After incubation, the cell media is replenished with the new warm cell media. Live imaging is carried out using a Nikon Eclipse Ti2 inverted widefield microscope equipped with a 10x objective, Kinetix Teledyne Photometrics sCMOS camera, and an Okolab stage top to maintain the 37\textdegree C temperature, 5\% CO$_2$, and humidity. The phase contrast images are acquired at different locations on the pattern at an interval of 15 minutes. At the same locations, the fluorescent images of the nuclei are taken every 1 h by exciting the Hoescht dye at 370 nm. The resolution of the images is 0.26 $\mu$m/pixel, and the field of view is 832 $\mu$m x 832 $\mu$m. The live imaging experiments are repeated three times independently.

\subsection{\label{matlab}Velocity fields and velocity correlation length}

After the frames are acquired, the phase contrast images are bleach-corrected with the histogram matching method using the Fiji \cite{fiji} image processing open source platform to remove any possible intensity fluctuation. The bleach-corrected phase contrast images are imported to PIVlab \cite{pivlab} in MATLAB. In PIVlab, the images are pre-processed by Contrast Limited Adaptive Histogram Equalization (CLAHE), auto contrast stretch, and background subtraction. The single pass direct cross correlation (DCC) algorithm is employed with an interrogation area of 200 pixels and a step 100 pixels. With this analysis, the velocity vectors are also calculated for the pixels that lay in blank spaces, with no cells. This is an issue until the cells reach confluency on the pattern. To remove such non-representative vectors from each frame, the magnitude of the velocity vectors is sorted in descending order. The cell-covered area is then estimated as the fractional cell coverage, calculated using the formula: (number of cells per frame * average cell spread area) / total frame area. The details on estimating the number of cells are provided in Section \ref{cell density}. The percentage of vectors with the lowest velocity equal to the percentage of empty surface area is then discarded. For example, if the area coverage is measured to be 75$\%$, the lowest 25$\%$ vectors are discarded. Then, the tangential (tangent to the ridges of the pattern) and radial components of the velocity vectors are extracted (see Fig. S1 in the supplementary information (SI)). The root mean square (rms) and average velocities for each frame are calculated as $v_{rms}=\sqrt{\frac{1}{n} \sum_{i} v{_i}{^2}}$ and $v_{avg}=\frac{1}{n} \sum_{i} v{_i}$, respectively, where $v{_i}$ and $n$ are the modulus of i-th vector and the total number of vectors in each frame, respectively. The interrogation area in PIVlab is chosen in order to ensure a good match between the rms velocities obtained from PIVlab and the velocities quantified by tracking the nuclei in the TrackMate plugin \cite{trackmate} in Fiji (see Fig. S2 in SI). For nuclei tracking, the video is taken with a 15 min interval between frames.

Further, the velocity-velocity spatial correlation function for each frame is calculated as follows:

\begin{equation}\label{corr_func}
C_{v_jv_j}(\delta r,t) =  \frac{\left\langle v_j(r+\delta r,t).v_j(r,t)\right\rangle}{\left\langle v_j(r,t)^2 \right\rangle}
\end{equation}

Here $r$, $\delta r$, and $t$ are the spatial coordinate, distance between vectors, and time, respectively. The index $j$ indicates the two main directions, tangential or radial, so $v_j$ can be either the tangential velocity $v_T$ or the radial velocity $v_R$. To calculate $C_{v_Tv_T}$, tangential velocity vectors located within concentric annular regions are selected, whereas for the $C_{v_Rv_R}$, radial velocity vectors are chosen from cone-shaped sectors intersecting the center of the frame. The correlation function is fitted to an exponential function to extract the velocity correlation length ($\xi_{v_{T}v_{T}}$: tangential correlation length and $\xi_{v_{R}v_{R}}$: radial correlation length). The velocity correlation lengths are averaged by a moving mean of 10 frames. 
    
\subsection{\label{cell density}Image processing of fluorescent images of nuclei}
After acquiring the frames of fluorescent images of the nuclei at an interval of 1 h, the frames are pre-processed by CLAHE and smoothing. Further, the autolocal thresholding (Phansalker method), median filtering, and watershed are performed in Fiji to segment the nuclei. All image processing steps are performed in Fiji \cite{fiji}. After segmenting the nuclei, the cell density is quantified for each frame. Furthermore, the large number density fluctuations are also calculated for each frame by following \cite{duclos2014perfect, turiv2020topology}: each frame is divided into smaller square interrogation windows of increasing size, and the mean number of cells $\left\langle N \right\rangle$ and the deviation $\Delta N$ from the mean value are measured.

\subsection{\label{nematic order}Nematic order}

To obtain the nematic order parameter in cells after confluency, the orientation angle field ($\theta$) from phase contrast images of cells is calculated using the OrientationJ plugin \cite{orientationj} in Fiji \cite{fiji}. This plugin measures the orientation for each pixel in a robust way using the structure tensor tool. In this, a Gaussian observation window is chosen with a local window size of 7.8 $\mu$m (= 30 pixels in our case). The justification for selecting this window size is given in Section I in SI. After extracting the orientation field, the spatial autocorrelation function of the nematic orientation is measured as follows \cite{li2019,copenhagen2021}:     

\begin{equation}
C_{2\theta}(r) =  
\left\langle \hat{n}_{2\theta}(r).\hat{n}_{2\theta}(0) \right\rangle
\end{equation}

where $\hat{n}_{2\theta} = (cos(2\theta), sin(2\theta))$ is the nematic director. The nematic correlation length is obtained as the intersection point of the fitted linear decay and the x-axis (see Fig. S4). With this method, the measured nematic correlation lengths of the system on PDL, FN, and xFN surface functionalization are 93$\pm$12 $\mu$m, 282$\pm$2 $\mu$m, and 280$\pm$1 $\mu$m, respectively. The highest value of correlation length, 282 $\mu$m, is then selected as a window size to calculate the nematic order parameter for each surface functionalization. The scalar nematic order parameter is calculated as twice the maximum eigenvalue of the 2D nematic order tensor $Q$ with:

\begin{equation}
Q =  \begin{bmatrix}
\left\langle \cos^2 \theta \right\rangle - \frac{1}{2} & \left\langle \cos \theta \sin \theta \right\rangle \\
\left\langle \cos \theta \sin \theta \right\rangle & \left\langle \sin^2 \theta \right\rangle - \frac{1}{2}
\end{bmatrix} 
\end{equation}

where brackets $\left\langle \right\rangle$ denote an average over a spatial window size of 282 $\mu$m. 

\subsection{AFM measurement}

Atomic force microscopy (AFM) is used to quantify the cell surface adhesion strength, following the procedure in \cite{nguyen2016}. One day after the cells are seeded on a plain substrate, the substrate is transferred to the AFM sample holder, and new warm media is added to the sample. The holder is placed and enclosed on a heating stage within the AFM chamber, whose temperature is maintained at 37\textdegree C. All experiments are carried out within an hour of the sample being removed from the incubator. The microscope NanoWizard 4 system (JPK, Bruker) used for this experiment is equipped with a Point Probe Plus-Non Contact/Soft Tapping Mode cantilever having a nominal force constant of 7.4 N/m. A Nikon Ti2-E inverted microscope with a 10× objective (Nikon ELWD S Plan Fluor, NA = 0.6), integrated with the AFM, is used to visualize the cells. The spring constant k and sensitivity S of the cantilever are determined by thermal noise, and the measurements are conducted in contact mode. Typically, the contact mode is used for scanning surface topography with a high gain frequency to maintain minimal contact between the tip and the sample. However, for the cell surface adhesion experiments, continuous contact between the cantilever and the cell is essential to drag the cell across the PDMS surface. To ensure this, a high scan speed of 76.05 $\mu$m/s and reduced gain frequency to 3 Hz are chosen, with a setpoint value of 0.3 V. A detailed mathematical description of the force measurement procedure is provided in Section II of the SI. In this section, the lateral force is calculated from the total interaction force recorded by the AFM cantilever tip \cite{Zhang2011}.

\section{\label{results}Results}

\subsection{Collective cell dynamics depends on cell-surface adhesion}

We study the dynamics of NIH 3T3 fibroblasts on topographically patterned substrates. Nematic liquid crystals are a good model to describe NIH-3T3 cells: in fact, these cells are elongated and tend to align with their neighbors; moreover they do not develop strong cell-cell junctions, so their motion is not fully constrained by the neighbors \cite{duclos2014perfect, duclos2017topological}. NIH-3T3 cells are grown on surfaces patterned with concentric ridges, creating arrays of topological defects with +1 and -1 topological charge, similar to those used in previous work \cite{Endresen2021, Kaiyrbekov2023}. The substrates are made of PDMS and coated with poly-D-lysine (PDL), fibronectin (FN), or covalently bonded fibronectin (xFN), as detailed in the Methods section. Poly-D-lysine is a synthetic polypeptide with a strong positive charge, while fibronectin is a protein of the extracellular matrix. Both PDL and FN are commonly used to enhance fibroblast growth in 2D cultures. FN can be either deposited on a surface or linked by covalent bonding (xFN) using gluteraldehyde \cite{kuddannaya,priyadarshani}. 

\begin{figure*}
\includegraphics[width=1\textwidth]{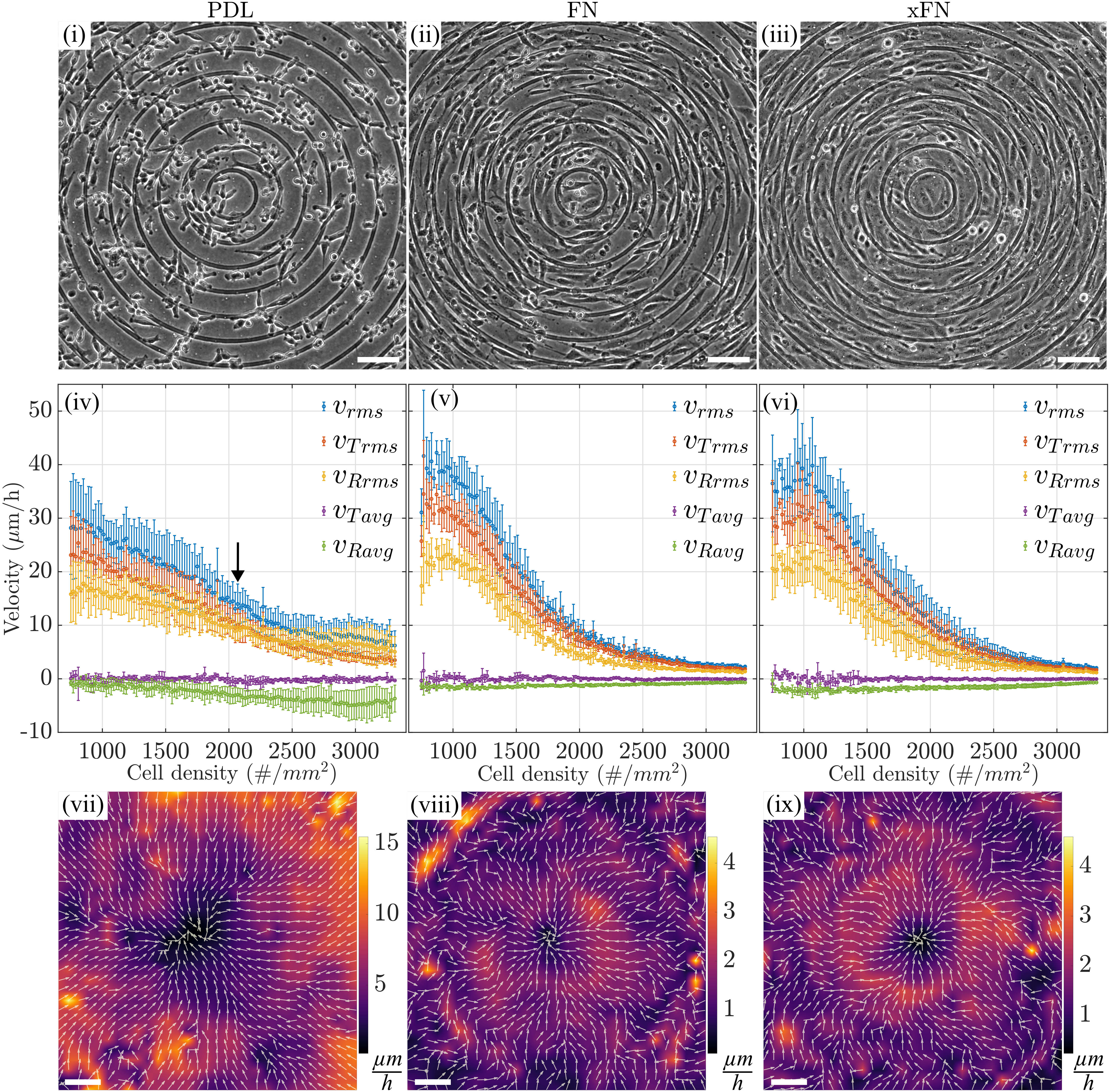}
\caption{\label{velocity} (i-iii) Phase contrast images of cells on (i) PDL, (ii) FN, and (iii) xFN surface functionalization at a cell density of 750 \#/$mm{^2}$ demonstrating different shapes of cells (Scale bar: 100 $\mu m$). (iv-vi) Velocities of cells with respect to cell density on surface functionalization of (iv) PDL, (v) FN, and (vi) xFN. Each graph contains root mean square (rms) velocity (in blue), rms tangential velocity (in red), rms radial velocity (in yellow), average tangential velocity (in violet), and average radial velocity (in green) with respect to the cell density. The data is shown as mean$\pm$sd for n = 4, 3, and 3 biological replicates for PDL, FN, and xFN, respectively; see Table I in SI for more details. The black arrow in (iv) indicates the crossover point between higher rms tangential and higher rms radial velocities. (vii-ix) Time-averaged velocity vectors (indicating direction) overlaid on a color map depicting the velocity magnitude for (vii) PDL, (viii) FN, and (ix) xFN. The velocity vectors were time-averaged over the last phase of the cell density from 2400 \#/mm$^2$ to 3000 \#/mm$^2$. Please note the differences in the velocity magnitude indicated by the color bar.}
\end{figure*}

Figures \ref{velocity}(i), (ii), and (iii) show the phase contrast images of cells on surfaces functionalized with PDL, FN, and xFN, respectively, at a cell density of 750 \#/$mm{^2}$. We can immediately observe that the cells on the PDL-coated surface have an irregular shape, whereas the cells on the FN and xFN surfaces show a characteristic elongated, spindle-like shape. 

Supplementary Videos 1, 2, and 3 demonstrate the cell dynamics on PDL, FN, and xFN, respectively. On PDL, in the initial phase, the cells are moving in every direction and dividing. As soon as they reach confluency and cover the whole surface, we observe a prominent inward motion of the cells, which slows down as the cell density increases. On the contrary, on FN, the cells follow the ridges at low density. As they proliferate, they tend to align with each other and move following the ridges in alternating clockwise-counterclockwise streams. As the cell density increases, the cell shape becomes isotropic in the inner rings. A similar behavior can also be seen in for cells on xFN, even though the cells tend to be larger and more elongated. The frames in the videos are analyzed using PIVlab in MATLAB to extract the velocity vectors. Thanks to fluorescent images of nuclei, we can quantify the cell density in every frame. A detailed description of extracting the velocities and the quantification of cell density is provided in the Methods sections \ref{matlab} and \ref{cell density}. The calculated cell densities increase with time as expected (Supplemental Fig. S5), and we can show that cell confluency occurs for PDL, FN, and xFN at cell densities of 2464$\pm$154, 1909$\pm$103, and 1528$\pm$115 \#/$mm{^2}$, respectively. The different numbers are consequences of the different cell sizes for each functionalization.

Figures \ref{velocity}(iv), (v), and (vi) show the velocity as a function of cell density for cells on PDL, FN, and xFN, respectively. The root mean square (rms) velocity decreases monotonically as the cell density increases, except for a few low-density data points on FN and xFN. This trend suggests that as cells divide, they occupy all the available surface, leading to a decrease in velocity. In the initial phase of the experiments, the rms tangential velocity (along the ridges) is larger than the rms radial velocity for every sample. Only on PDL, at a particular cell density just below confluency, the rms tangential velocity becomes equal to the rms radial velocity (shown by the black arrow in Fig. \ref{velocity}(iv)). On the other hand, on FN and xFN, the rms tangential velocity keeps dominating the rms radial velocity. For each surface treatment, the average tangential velocity remains zero during the whole experiment, suggesting that the cells move clockwise or anticlockwise with the same probability. However, the negative average radial velocity indicates that the motion of the cells on the pattern is always directed toward the +1 defect.  On FN and xFN, cells have a small, essentially constant inward motion toward the +1 defect with drift speed around 1-2 $\mu$m/hour, which eventually is suppressed at high density when the cells' velocity decreases. On PDL, the situation is more complex: the average radial velocity is small at low density, then increases to about 5 $\mu$m/hour. We analyze the average radial velocity for different annular regions. It is observed that the peak in the average radial velocity is reached at different densities (defined as the total number of cells/total frame area) in different regions (see Fig. S6). This shift in the peak value suggests that cells first fill the inner region, followed by the adjacent outer regions.       
 
Figures \ref{velocity}(vii-ix) show the time-averaged velocity vectors for each group overlayed on top of the color map of the magnitude of the velocity of cells at very high density (data are taken as a mean of the frames obtained for cell density 2400 to 3000 \#/mm$^2$). The velocity vectors are pointing inward with a higher magnitude in the PDL case than in the FN and xFN. This result also confirms the observation of cell movement in the videos.   

\subsection{Quantification of cell-surface adhesion strength}

\begin{figure}
\includegraphics[width=0.45\textwidth]{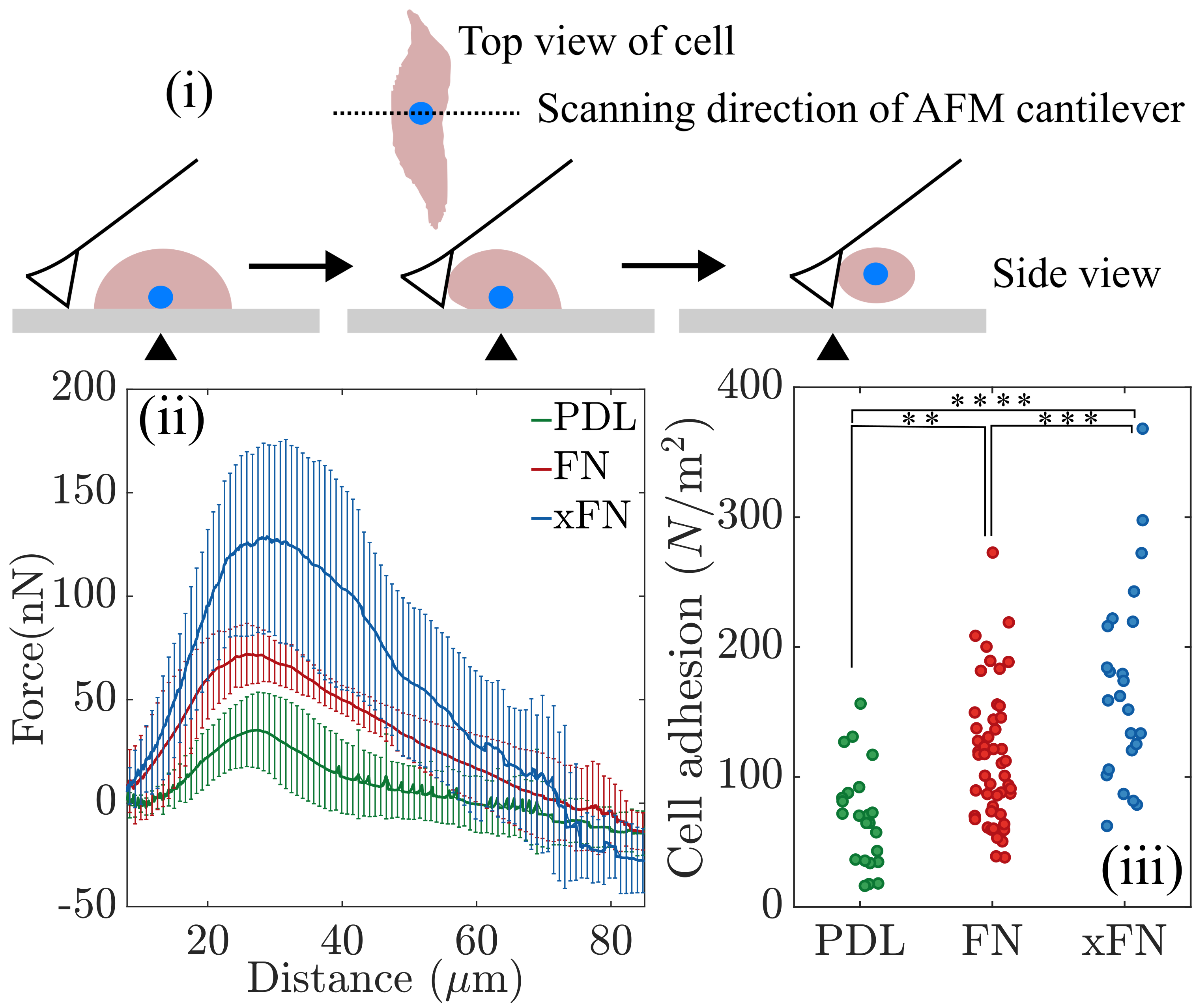}
\caption{\label{AFM} (i) Schematic illustration of an AFM cantilever tip dragging a cell perpendicular to its main axis, (following the method in \cite{nguyen2016}) (ii) Lateral force vs distance curve obtained by dragging the AFM cantilever tip perpendicular to the cell orientation for PDL, FN, and xFN. The experiments were repeated n=23, n=49, n=24 times for PDL, FN, and xFN, respectively, on different days, and the force is plotted as mean$\pm$sd. (iii) The surface adhesion of the cells on the coatings is calculated by dividing the maximum force by the cell spread area. The P value measured in the post hoc test shows a clear difference between the three different substrate coatings ($**$ pvalue=0.0044, $*$$*$$*$ pvalue=0.0006, and $*$$*$$*$$*$ pvalue$<$0.0001).}
\end{figure}

The cell-surface adhesion strength is quantified using AFM to determine the effect of different surface functionalization, following the method introduced by Nguyen et al.\cite{nguyen2016}. In this method, the AFM tip is used to detach the cells from the substrates after impacting the side of the cells and dragging the cells along the surface (Fig. \ref{AFM}(i)). The adhesion is calculated from the force exerted on the cantilever from its impact with the cell until the detachment of the cell. The AFM cantilever tip should be moving in a direction perpendicular to the cell orientation at the time of impact and the impact should occur near the nucleus. Figure \ref{AFM}(ii) shows the obtained lateral force vs. distance graphs of the cells on the surface functionalization of PDL, FN, and xFN. The detailed description of the lateral force measurement is given in Section II of the SI. Cell-surface adhesion strength is calculated as the ratio of the maximum force to the cell spread area. The results are plotted in Fig. \ref{AFM}(iii), where each point corresponds to a cell, and the cells are grouped according to the surface functionalization. The results reveal that cells on the PDL-coated surface exhibit significantly lower adhesion strength compared to those on other coatings. In fact, the data may overestimate adhesion on PDL, as the least adhering cells detach from the surface during the transfer of samples to the AFM stage. As a result, the measured cells on the PDL are those with the strongest adhesion. The data also show a significant difference in adhesion strength between the FN and xFN treatments. 

The AFM data corroborate our hypothesis that PDL provides the weakest adhesion. This is consistent with various observations. First, the cell shape is more irregular on PDL. Moreover, we can observe cell attachment just after seeding on the surface functionalization of PDL, FN, and xFN. Figure S7 shows phase contrast images of the attachment of the cells 35 minutes after seeding. In the case of PDL, most cells are still in a spherical shape, whereas in the case of FN and xFN the cells start spreading. It suggests that the cells take longer to spread and attach to the substrate on the PDL-coated surface. Finally, it is consistent with our data on cell velocity, because a lower adhesion on PDL may facilitate a greater number of cells crossing the height barrier of the ridges and favor inward motion. 

\subsection{Cells' nematic order parameter}

\begin{figure}[ht]
\includegraphics[width=0.48\textwidth]{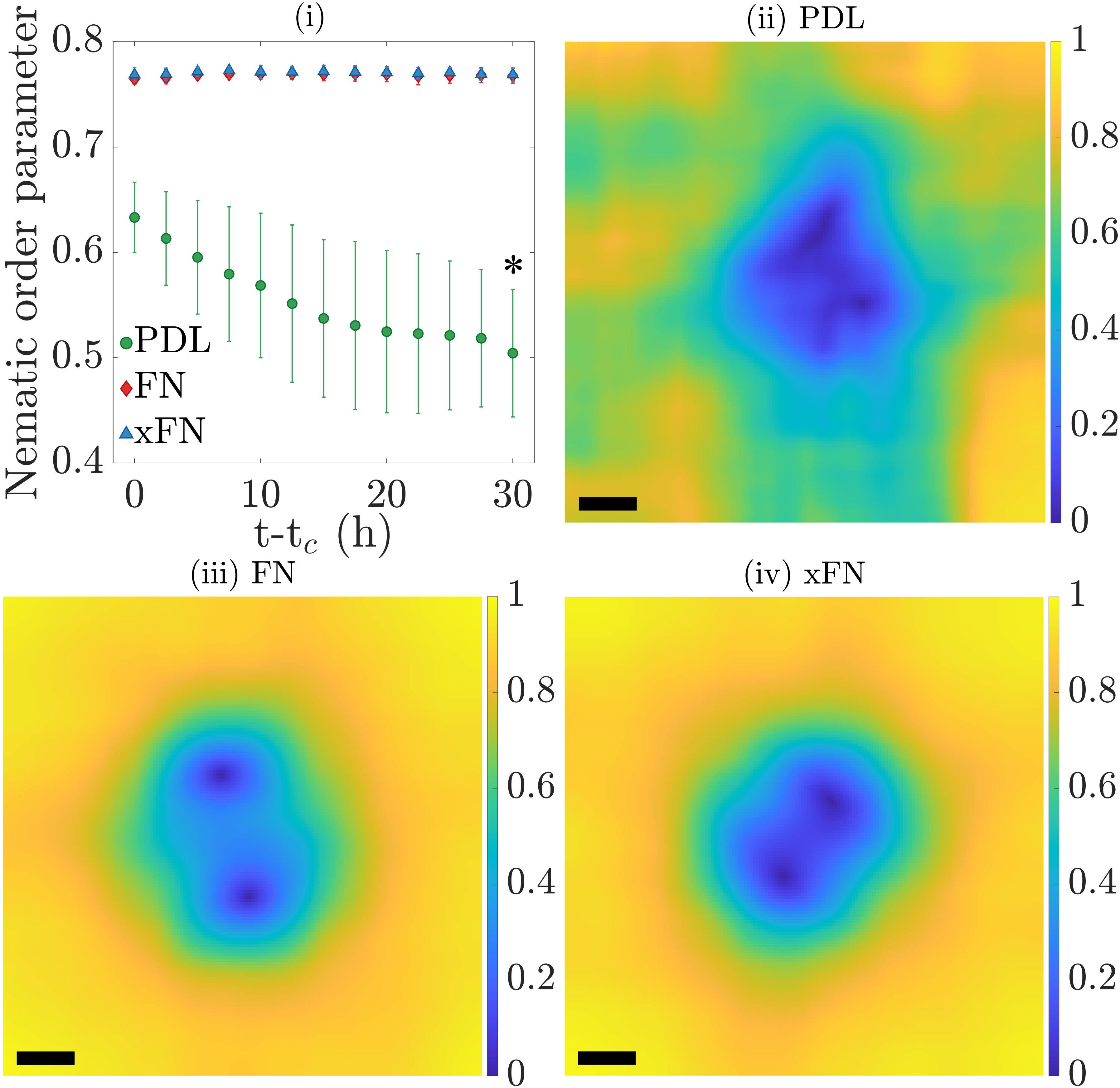}
\caption{\label{spatial_OP} (i) Order parameter (mean$\pm$sd, n = 4, 3, and 3 biological replicates for PDL, FN, and xFN, respectively; see Table I in SI for more details.) of cells on PDL, FN, and xFN surface functionalization as a function of time after confluency (t$_c$: time of confluency). $*$ p $\textless$ 0.0042 (Bonferroni-corrected), compared with order parameter the t$_c$ using paired t-test. Spatial nematic order parameter of cells on (ii) PDL, (iii) FN, and (iv) xFN surface functionalization at a cell density of 2500 \#/mm$^2$ (Scale bar: 100 $\mu$m).} 
\end{figure}

We calculate the local nematic order parameter, which identifies the degree of order in the system, for the different surface treatments. The detailed description of how we obtain the order parameter is discussed in \ref{nematic order}. Figure \ref{spatial_OP}(i) shows the calculated scalar nematic order parameter as a function of time post-confluency for each surface functionalization. Note that we calculate this quantity only after confluency, i.e., when there are no empty spaces without cells. Cells on PDL tend to lose their order after confluency, whereas cells on FN- and xFN-coated surfaces maintain their aligned position along the ridges of the pattern. In the case of PDL, the order parameter at each time point is compared with the order parameter at confluency using a paired t-test. The nematic order has also been analyzed at a cell density of 2500 \#/mm$^2$, where all samples are above confluency. The spatial distribution of the order parameter reveals a diffused disordered region for PDL around the center of the +1 defect and two distinct low-order parameter regions that may correspond to two +1/2 topological defects for FN and xFN, as shown in Fig. \ref{spatial_OP}. Interestingly, two +1/2 topological defects is the equilibrium configuration of nematic liquid crystals confined within a circular geometry, where two +1/2 defects are known to form as a result of boundary conditions and topological constraints \cite{de1993physics}, despite the fact that cells are out-of-equilibrium systems. This finding is consistent with the result reported by Duclos et al., \cite{duclos2017topological} for cells confined to a circular patch. For confined cells, it is shown that the distance of the defects from the center of the patch, $d$, scales with the radius of the patch $R_p$, with $d=0.67 R_p$. In our experiment, the distance between the two defects for FN and xFN is 171$\pm$9 $\mu$m and 175$\pm$8 $\mu$m, respectively. Using the same ratio between radius and defect distance (0.67) reported in ref. \cite{duclos2017topological}, the effective radius of confinement is calculated to be 131 $\mu$m, which is very similar to the radius of the second innermost rings of the topographical pattern. In our previous work \cite{Endresen2021} we observe that if the inner ring radius is much smaller than the orientation correlation length the cells form defects outside the inner ring.

\subsection{Velocity correlation length and large number density fluctuations}

\begin{figure}[ht]
\includegraphics[width=0.48\textwidth]{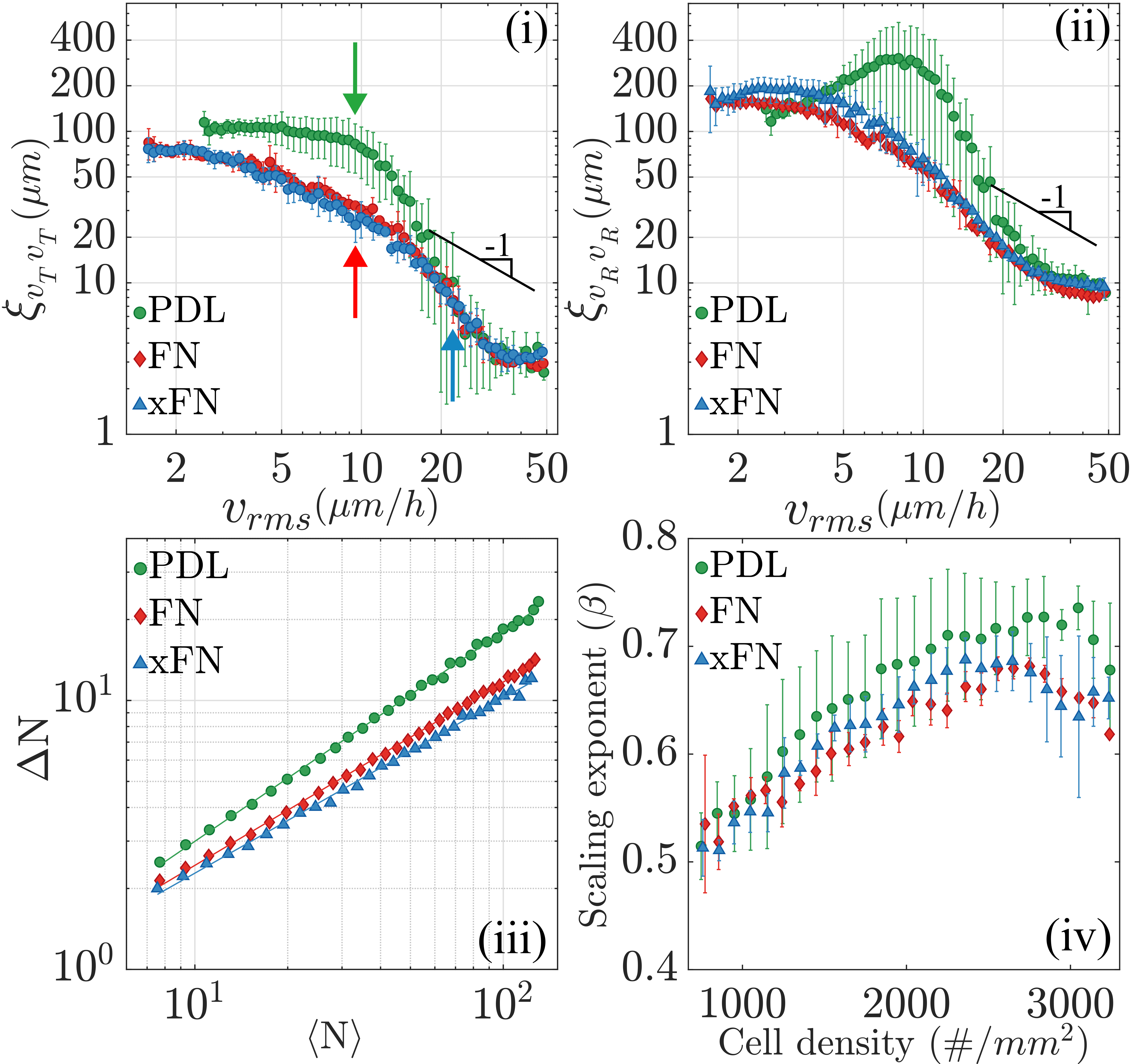}
\caption{\label{density fluctuation} (i) and (ii) tangential and radial velocities correlation length with respect to the rms velocity, respectively (log-log plot, mean$\pm$sd, n = 4, 3, and 3 biological replicates for PDL, FN, and xFN, respectively; see Table I in SI for more details). The green, red, and blue arrows for PDL, FN, and xFN, respectively, in (i) correspond to the velocity at cell confluency, (iii) Large number density fluctuations $\Delta N$ with respect to mean number of cells $\left\langle N \right\rangle$ (log-log plot) for PDL, FN, and xFN surface functionalization at a cell density of $\sim$ 2700 cells$/$mm$^2$. (iv) Scaling exponent ($\beta$) of density fluctuations obtained after fitting the $\Delta N$$\sim$$\left\langle N \right\rangle$$^\beta$ (mean$\pm$sd, n = 4, 3, and 3 biological replicates for PDL, FN, and xFN, respectively; see Table I in SI for more details) of cells on PDL, FN, and xFN surface functionalization with time after the confluency.} 
\end{figure}

To better understand the collective motion of the cells, we calculate tangential velocity correlation length ($\xi_{v_{_T}v_{_T}}$) and radial velocity correlation length {$\xi_{v_{_R}v_{_R}}$}, using equation (\ref{corr_func}). Figure S8 in the SI shows a typical graph of $C_{v_{_T}v_{_T}}$, the spatial correlation function of the tangential velocity, for one particular time point, with an exponential fit. The calculated $\xi_{v_{_T}v_{_T}}$ and $\xi_{v_{_R}v_{_R}}$ are plotted with respect to $v_{rms}$ as shown in Fig. \ref{density fluctuation}(i) and (ii), respectively, following the approach in \cite{garcia2015}. Both tangential and radial velocity correlation lengths increase from a few $\mu m$ to the order of several cell sizes as the $v_{rms}$ decreases.  $\xi_{v_{_R}v_{_R}}$ is always higher than $\xi_{v_{_T}v_{_T}}$ for each group, indicating that the collective radial motion of cells is more highly correlated than the tangential motion. The small value of $\xi_{v_{_T}v_{_T}}$ is consistent with the observation that NIH 3T3 tends to change their polarity often and spontaneously invert their direction of motion \cite{duclos2014perfect}, which results in a shorter correlation length. 

Initially, in the higher $v_{rms}$ regime, $\xi_{v_{_T}v_{_T}}$ and $\xi_{v_{_R}v_{_R}}$ are small and approximately equal across all surface functionalizations. As cells grow and slow down, the $\xi_{v_{_T}v_{_T}}$ for PDL are higher than those for FN and xFN. At early times, where $v_{rms}$ is higher, both $\xi_{v_{_T}v_{_T}}$ and $\xi_{v_{_R}v_{_R}}$ scale inversely with $v_{rms}$, exhibiting a slope smaller than -1. According to the model presented by Garcia et al \cite{garcia2015}, a slope of exactly -1 indicates a regime dominated by cell-surface adhesion. This adhesion-dominated regime is also apparent in our results. Furthermore, the slope magnitude is larger for PDL than for FN and xFN, likely due to differences in surface chemistry arising from surface functionalization. In contrast, at lower $v_{rms}$ values, the $\xi_{v_{_T}v_{_T}}$ and $\xi_{v_{_R}v_{_R}}$ begin to approach a constant value for FN and xFN, while $\xi_{v_{_R}v_{_R}}$ for PDL exhibits a distinct decline that coincides with the onset of complete surface coverage by the cells.

In addition, we characterize the system by quantifying the large number density fluctuations $\Delta N$. $\Delta N$ follows the relationship as $\Delta N$$\sim$$\left\langle N \right\rangle$$^\beta$, where $N$ is the number density and  $\beta$ is the scaling exponent: according to theory, $\beta$ is 0.5 in passive systems at equilibrium and 1 (theoretical limit) in active nematics \cite{ramaswamy2003}. The details of the estimation of $\Delta N$ and $\left\langle N \right\rangle$ are provided in Sec. \ref{cell density}. Figure \ref{density fluctuation}(iii) shows the linear relationship between $\Delta N$ and $\left\langle N \right\rangle$ for PDL, FN, and xFN at a cell density of $\sim$ 2700 cells$/$mm$^2$. The $\beta$ is extracted from the linear fit for each frame and plotted as a function of cell density, as shown in Fig. \ref{density fluctuation}(iv). Initially, the values of $\beta$ are close to 0.5 when the cells are at low density. For each surface functionalization, $\beta$ increases with the cell density, consistently with previous studies \cite{narayan2007, zhang2010}. At higher cell densities, $\beta$ greater than 0.5 suggests that cells behave as active nematics. The maximum average $\beta$ attains a value of 0.74, approaching the characteristics value of 0.8 observed in Toner and Tu phase, the homogenous ordered phase of the Viscek class \cite{ginelli2016}. While the scaling exponent appears slightly larger for cells on PDL, this behavior does not show a strong dependence on surface coating. 

\section{\label{results}Discussion}
We demonstrate distinct collective cell dynamics of NIH 3T3 fibroblast cells by varying cell–surface adhesion strength at integer topological defects. Fibroblast cells are elongated and lack strong cell-cell junctions, unlike epithelial cells, making their motion less constrained by neighbor cells \cite{duclos2014perfect, duclos2017topological}. Therefore, NIH 3T3 constitutes a good model of nematic liquid crystals. PDL and FN are chosen as extracellular matrices for surface functionalization. PDL is a chemically synthesized polypeptide, whereas FN is a well-studied fibrillar ECM protein in the literature. Collective cell migration guided by FN has been demonstrated in wound healing, tumor tissues, and neural crest \cite{humphries1989, gopal2017, martinson2023}. In our experiments, we observe two distinct modes of collective cell dynamics for cells grown on PDL- and FN-coated surfaces, respectively. Cells exhibit collective inward radial motion on PDL-coated topographical patterns (weaker cell–substrate adhesion), while cells on FN- and xFN-coated patterns (stronger adhesion) move predominantly tangentially along the ridges of the pattern. Interestingly, Ravasio et al. \cite{ravasio2015} demonstrate that MDCK colonies on flat substrates exhibit rotational or outward migration depending on FN concentration, governed by the balance between cell–cell cohesion and cell–substrate adhesion: weak substrate adhesion favors rotation, whereas stronger adhesion enables colony expansion. This difference likely arises from two key factors: cell type and substrate geometry. In our work, cells on FN- and xFN-coated surfaces show similar dynamic characteristics, including tangential and radial rms, velocity correlation lengths, and large number density fluctuations, despite their markedly different adhesion strength and the morphological differences of cells on the two substrates. For example, the onset of confluency occurs at different cell densities: 1909$\pm$103 \#/$mm{^2}$ for FN, and 1528$\pm$115 \#/$mm{^2}$ for xFN, indicating a much larger spread area of cells on xFN than on FN. 

In the literature, it has been shown that the cells move towards the +1 topological defect core irrespective of defect types (asters, spirals, and targets) \cite{zhao2025integer}. In the present work, we show that cells migrate to the defects irrespective of the surface treatment, but the behavior is enhanced for PDL. 

We characterize the nematic order and find, as expected, high order on FN- and xFN-coated surfaces, but low order on PDL. The equivalence between FN and xFN suggests that the order does not directly depend on adhesion. We show that the strongly adherent surfaces tend to suppress +1 defects and split them into two +1/2 defects, while the PDL-coated surfaces exhibit a single, almost circular low order parameter region consistent with the presence of a +1 defect. This suggests that by modulating the surface coating and arresting cell growth at defined time points, it becomes possible to engineer tissue organization incorporating integer defects.

We measure the velocity correlation length and interpret our data using the results reported in Garcia et al \cite{garcia2015}, where the scaling of the velocity correlation length results from the interplay of cell-substrate and cell-cell friction. Garcia et al. show that $\xi_{vv}$ $\propto$ 1/$v_{rms}$ for high $v_{rms}$ regime in the case of NIH 3T3 cells. Our observations show a slope smaller than -1 for high $v_{rms}$ values.

Finally, we determine the scaling exponent ($\beta$) of the large number density fluctuations as a function of cell density and find it mostly independent of adhesion strength and consistent with earlier observations in other active nematic systems, such as millimeter-long vibrated copper rods \cite{narayan2007} and micron-sized bacteria \cite{zhang2010}. The characteristic length scale of our present system lies between that of the vibrated rods and the bacteria, indicating that the increase of the scaling exponent $\beta$ with density could be a universal feature of active nematics across very different length scales.

In conclusion, we show that the collective dynamics of NIH 3T3 fibroblasts at integer topological defects is greatly affected by the surface coating. This effect arises not only due to adhesion strength, but also depends on the chemical nature of cell–substrate interactions. By functionalizing the surface with PDL, FN, and xFN, we observe distinct behaviors, such as a strong inward radial motion on PDL-coated surfaces and tangential motion on FN-coated ones. Despite marked differences in adhesion strength, FN and xFN coatings give cell layers similar nematic order, velocity correlation length, and large number density fluctuations. This may indicate that surface chemistry, rather than absolute adhesion force, governs collective cell behavior. These findings highlight how controlled surface chemistry and topographical confinement can be used to direct active organization in living cell monolayers and engineered tissues.

\begin{acknowledgments}
PA and FS would like to thank Prof Nir S. Gov for his suggestion on the velocity correlation length. FS acknowledges funding from Novo Nordisk Fonden, recruit grant NNF21OC0065453.
\end{acknowledgments}

\bibliography{references}

\end{document}


\preprint{APS/123-QED}
    \section*{Supplementary Information}
    \title{Surface functionalization modulates collective cell behavior at integer topological defects}
    
    
    \author{Prasoon Awasthi}
     \email{praw@sdu.dk}
    \author{Aniruddh Murali}
    \author{Ellen Juel Pørtner}
    \author{Adam Cohen Simonsen}
    \author{Francesca Serra}
     \email{serra@sdu.dk}
     \affiliation{
     Department of Physics, Chemistry and Pharmacy,  University of Southern Denmark.
    }
    
    \maketitle
    
    
    \begin{table}[h!]
    \centering
    \caption{Number of data points used in the calculations.}
    \begin{tabular}{|c|c|c|c|c|c|}
    \hline
    \parbox[c][1.2cm][c]{3cm}{\centering Figure} &
    \parbox[c][1.2cm][c]{4cm}{\centering Surface functionalization} &
    1st & 2nd & 3rd & 4th \\
    \hline
    1(iv) & PDL & 8 & 9 & 4 & 6 \\
    1(v) & FN & 3 & 10 & 8 & - \\
    1(vi) & xFN & 8 & 8 & 3 & - \\
    3(i) & PDL & 8 & 9 & 4 & 6 \\
    3(i) & FN & 3 & 10 & 8 & - \\
    3(i) & xFN & 8 & 8 & 3 & - \\
    4(i)\,(ii) & PDL & 8 & 9 & 4 & 6 \\
    4(i)\,(ii) & FN & 3 & 10 & 8 & - \\
    4(i)\,(ii) & xFN & 8 & 8 & 3 & - \\
    4(iv) & PDL & 8 & 9 & 4 & 6 \\
    4(iv) & FN & 3 & 10 & 8 & - \\
    4(iv) & xFN & 8 & 8 & 3 & - \\
    \hline
    \end{tabular}
    \end{table}

    \begin{figure}
    \includegraphics[width=0.5\textwidth]{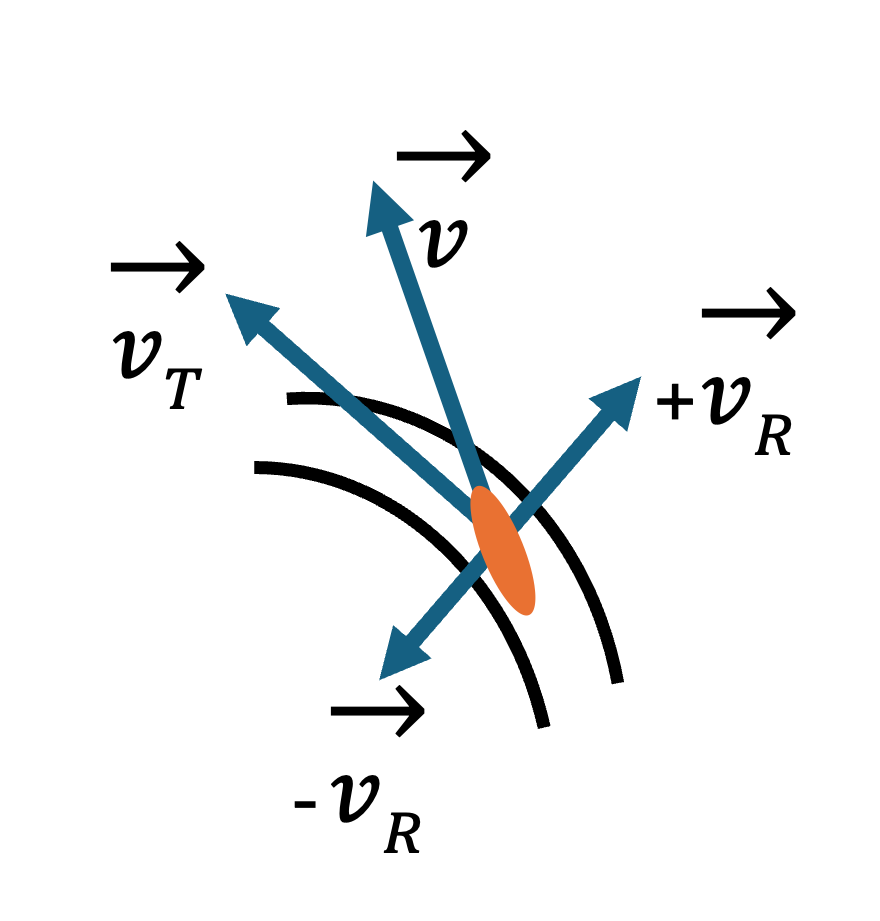}
    \caption{\label{schematic_velcity_vectors} Schematic shows velocity vector and its tangential (tangent to the ridges of the pattern) and radial component.}
    \end{figure}

    \begin{figure}
    \includegraphics[width=0.7\textwidth]{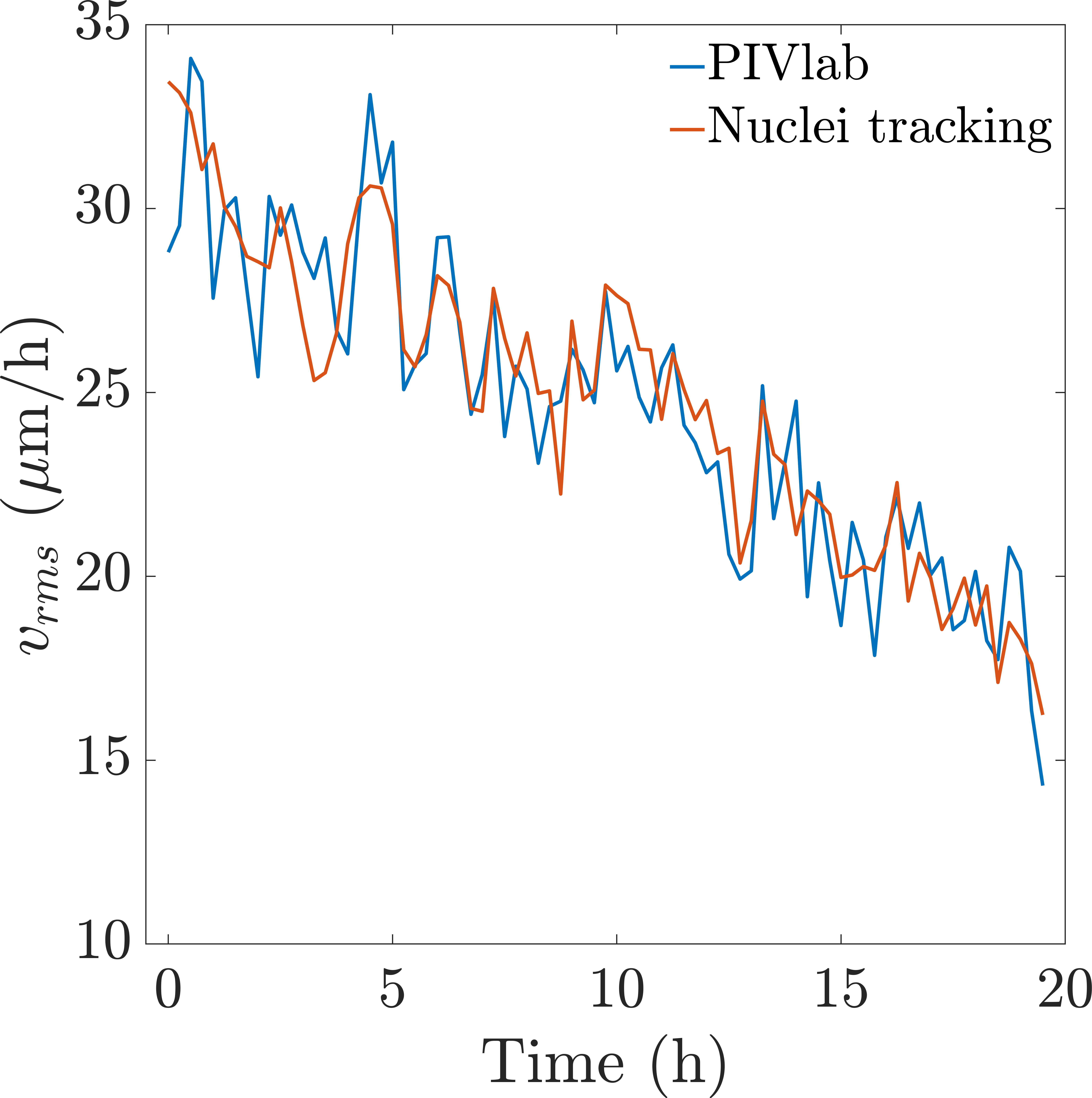}
    \caption{\label{PIVvsTrackMate} Comparing the root mean square velocity ($v_{rms}$) calculated using PIVlab and nuclei tracking. In PIVlab, the interrogation area was 200 pixel. For tracking the nuclei, the video had 15 min frame interval.}
    \end{figure}
    
    \section{\label{Jutification}Orientation field}
\begin{figure}
    \includegraphics[width=1\textwidth]{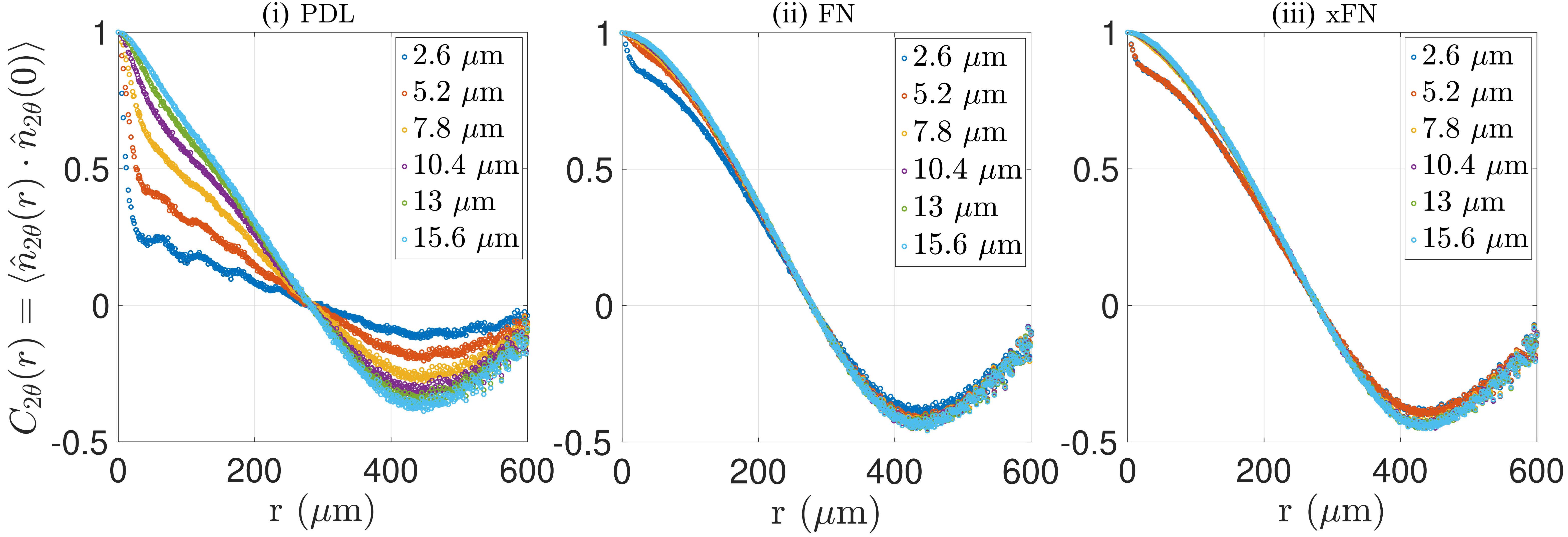}
    \caption{\label{30px}  The graphs (i), (ii), and (iii) are showing the spatial autocorrelation function of the nematic orientation for different local window sizes for PDL, FN, and xFN surface functionalization, respectively.}
    \end{figure}

    To identify the best possible local window size in the OrientationJ plugin to get the orientation field of cells, the local window size was varied. For each window size, the spatial autocorrelation function of the nematic orientation was calculated as shown in Fig. S\ref{30px}. As the window size increases, the autocorrelation plateaus after 7.8 $\mu$m in case of FN and xFN and 13 $\mu$m for PDL. So, the window size 7.8 $\mu$m was finalized to obtain the orientation field, and this size is also roughly the thickness of cells. In the OreintationJ plugin, it is also recommended to have the window size roughly close to the thickness of the structure.
    
    \clearpage
    \section{AFM measurement}
    
    The cell adhesion was measured based on the technique reported by Nguyen et al., 2016 \cite{nguyen2016}. During scanning in the AFM experiment, the force applied by the tip was determined by Hooke's law ($ F=k\cdot d$), in which the cantilever is modeled as a spring with a spring constant $k$ and a deflection $d$. This method allows for a direct calculation of the applied force based on the vertical displacement of the cantilever. However, when the AFM probe encounters a large feature, the exerted force acts perpendicular to the contact plane. In such scenarios, the sample exerts a force on the cantilever, causing it to bend more as it resists detachment. The total applied force of the probe on the cell can then be decoupled into the force exerted perpendicular to the surface and the force exerted along the surface (lateral force). The lateral force is responsible for detaching the cell. Consequently, the lateral force $F_{lat}$ used to quantify cell detachment was calculated following the method described in \cite{Zhang2011, nguyen2016}     
    
    \begin{gather}
        F_{lat}=kSV_{tot}\sin{\left(\Phi+\theta\right)}\cos{\theta'} \label{F_lat}
    \end{gather}
    
    where $k$ is the spring constant of the cantilever (nN/nm), $S$ is the sensitivity of the cantilever (nm/V), $V_{tot}$ is the electrical output from the photodiode relative to the measured vertical deflection of the reflected laser beam, $\Phi$ is the fixed cantilever orientation, and $\theta$ is the fixed angle determined by the probe geometry.
    To calculate the Eq.\ref{F_lat}, the angle $\theta'$ needs to be calculated, which represents the angle at which the force is applied by the cantilever on the cell. Before that, $\Phi'$ is calculated for the system, which represents the deflection of the cantilever due to deformation.
    \begin{gather}
        \Phi'=2\arctan\left[ \frac{L-\sqrt{\left(V_{tot}S\right)^2+\left(L\cos{\Phi}\right)^2}}{V_{tot}S+L\sin{\Phi}}\right]\label{phi'}
    \end{gather}
    where L is the length of the cantilever lever. With this, $\theta'$ can be calculated as follows:
    \begin{gather}
        \theta'=\Phi+\theta-\Phi'\label{theta'}\\
        \theta'=\Phi+\theta-2\arctan\left[ \frac{L-\sqrt{\left(V_{tot}S\right)^2+\left(L\cos{\Phi}\right)^2}}{V_{tot}S+L\sin{\Phi}}\right]\label{theta'_final}
    \end{gather}
    
    Finally, substituting the value of $\theta'$ from Eq.\ref{theta'_final} in Eq.\ref{F_lat}, $F_{lat}$ is given by:
    \begin{gather}
        F_{lat}=kSV_{tot}\sin{\left(\Phi+\theta\right)}* \cos{\Biggl\{\Phi+\theta-2\arctan\left[ \frac{L-\sqrt{\left(V_{tot}S\right)^2+\left(L\cos{\Phi}\right)^2}}{V_{tot}S+L\sin{\Phi}}\right]\Biggl\}} \label{F_lat_final}
    \end{gather}

    \begin{figure}
    \includegraphics[width=1\textwidth]{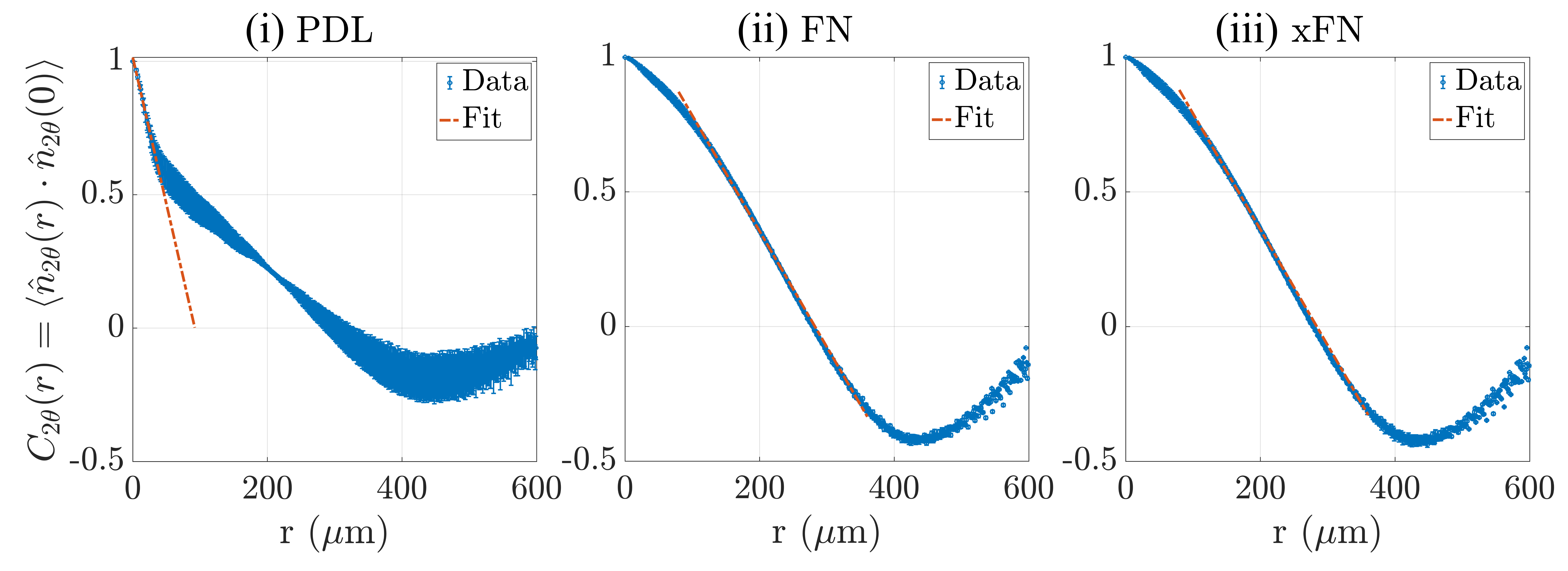}
    \caption{\label{nematiclength}  The graphs (i), (ii), and (iii) demonstrate the spatial autocorrelation function of the nematic orientation (mean$\pm$sd) for PDL, FN, and xFN surface functionalization, respectively. A linear decay was fit to measure the nematic correlation length.}
    \end{figure}

    \begin{figure}[t!]
    \includegraphics[width=0.65\textwidth]{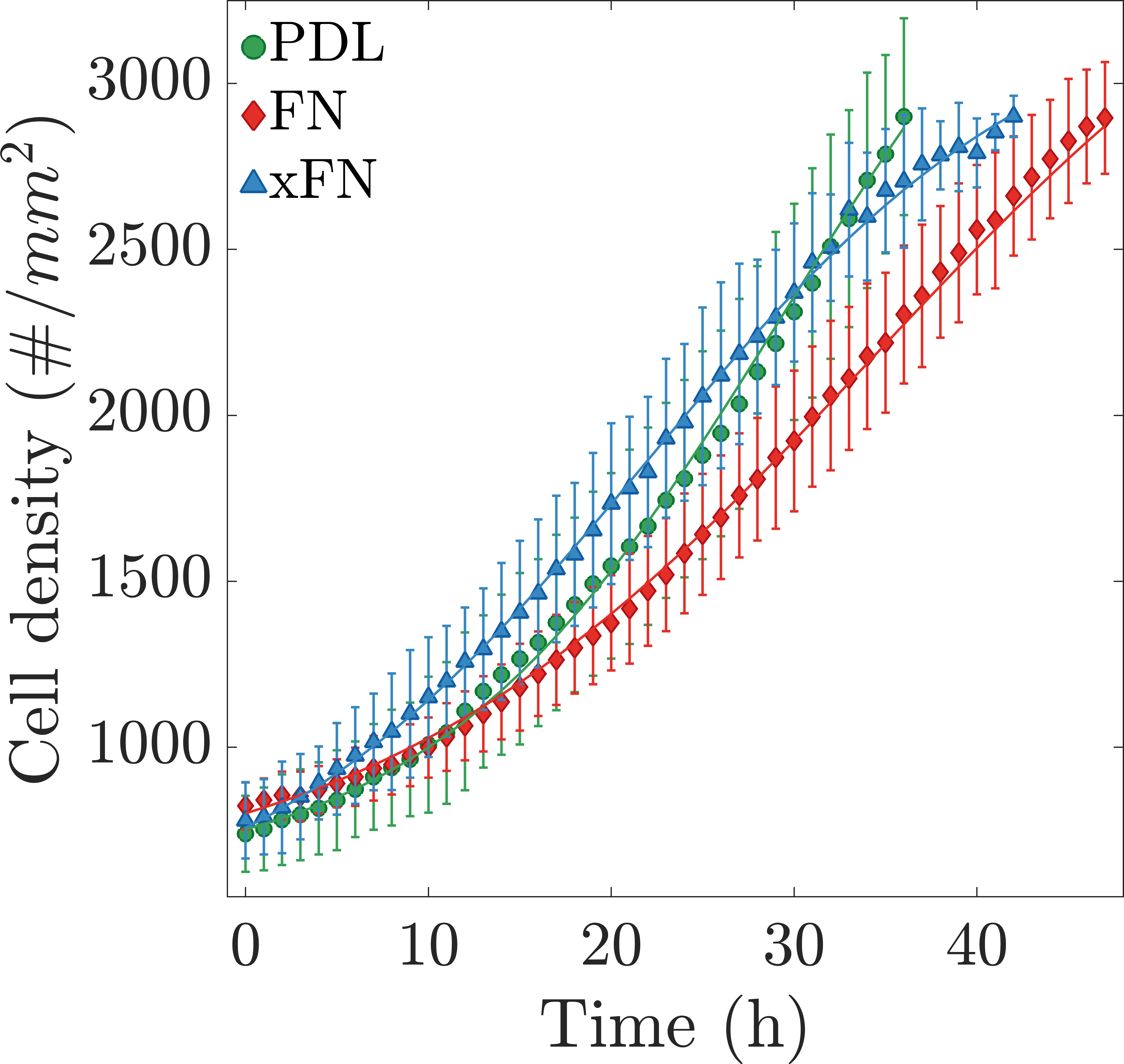}
    \caption{\label{celldensity_time}  This plot shows the cell density (number of cells per mm$^2$) (mean$\pm$sd, 3$\leq$n$\leq$10, n: number of observations in one experiment) with time for each surface functionalization of one experiment. The symbols and solid lines are experimental and average data fitting with a sigmoidal function, respectively. The green color circles, red color squares, and blue color triangles correspond to poly-D-lysine (PDL), fibronectin (FN), and covalently bonded fibronectin (xFN) surface functionalization, respectively. Three independent experiments were repeated, and the average cell densities were fitted, which gives the mean growth rates 0.1 h$^{-1}$, 0.08 h$^{-1}$, and 0.1 h$^{-1}$ for PDL, FN, and xFN surface functionalization, respectively.}
    \end{figure}

    \begin{figure}
    \includegraphics[width=1\textwidth]{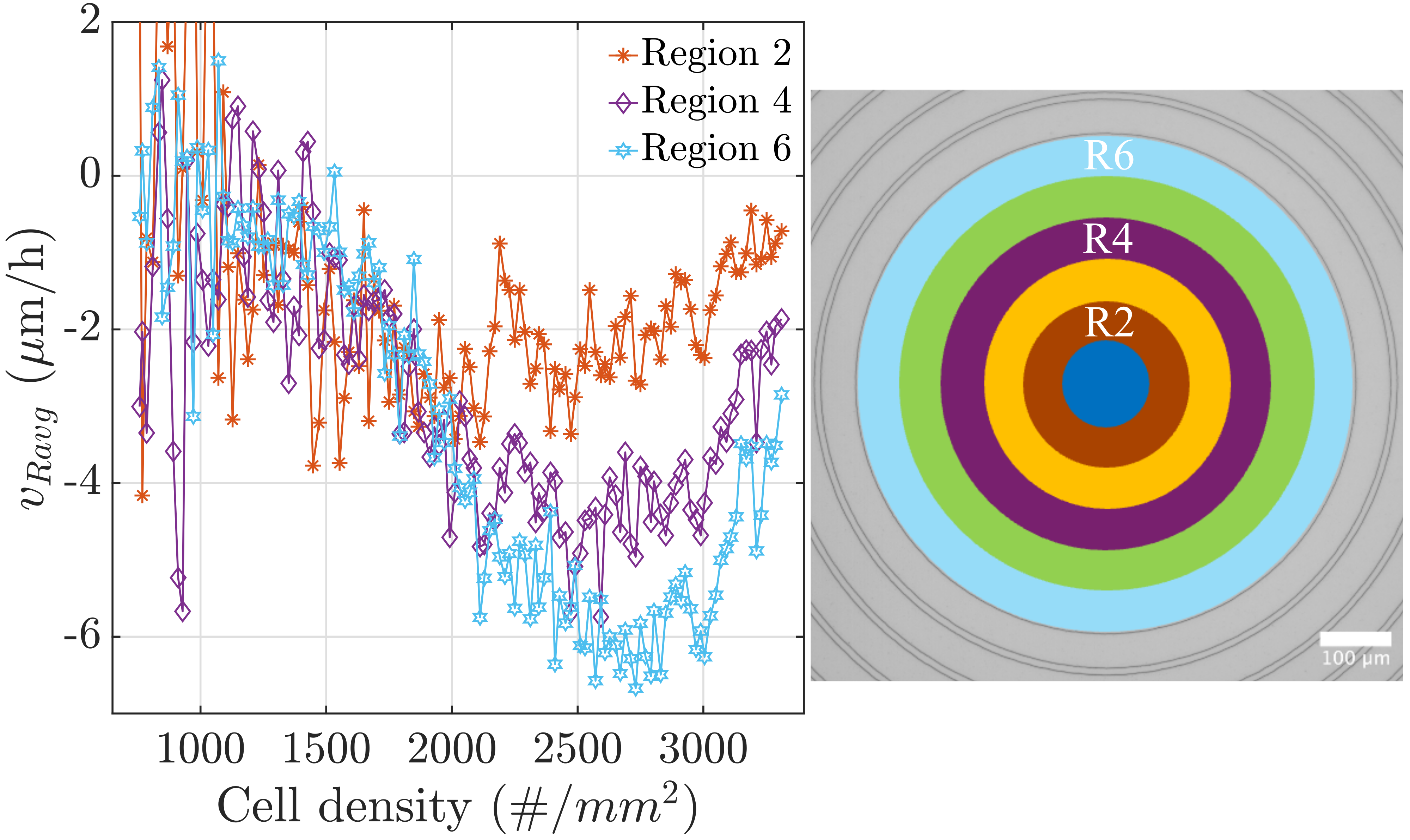}
    \caption{\label{regionwisevelocity} The mean $v_{Ravg}$ (9 observations in one experiment) with respect to cell density in annular regions 2, 4, and 6 for the PDL-coated surface (selected only three regions for better visualization).}
    \end{figure}
    
    \begin{figure}
    \includegraphics[width=1\textwidth]{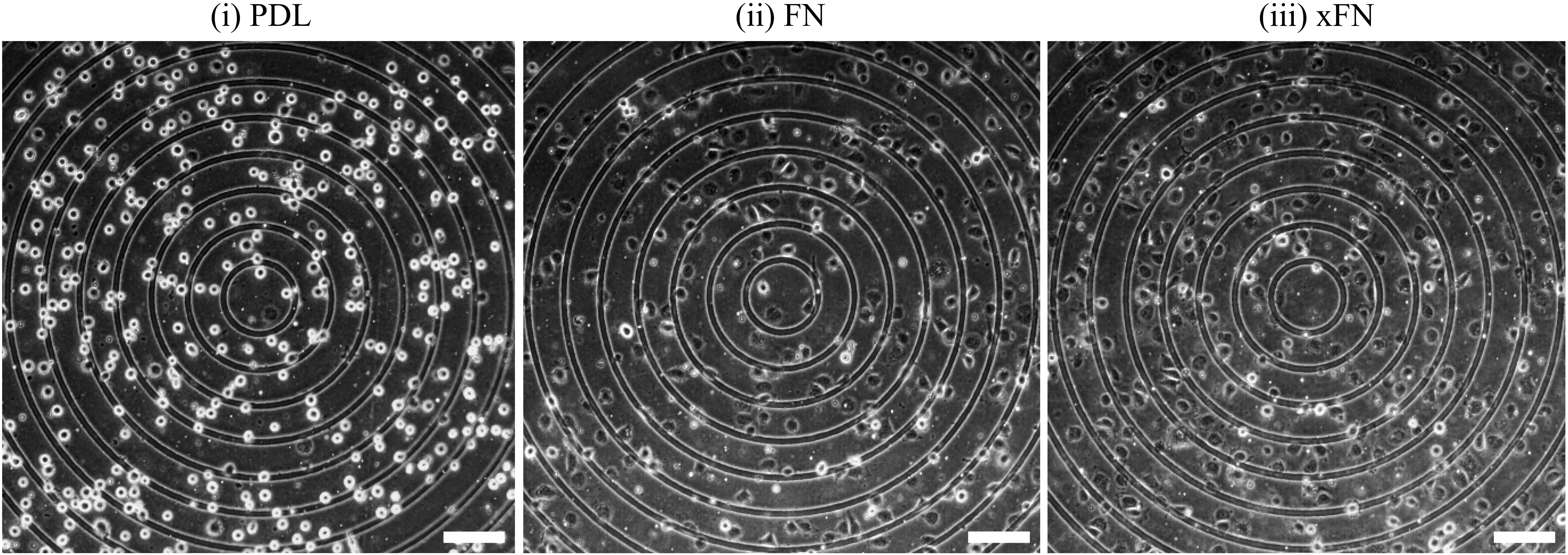}
    \caption{\label{initialattachment}  Phase contrast images of cells on surface functionalization of (i) PDL, (ii) FN, and (iii) xFN, respectively, after 35 minutes of cell seeding at a cell density of 200 \#/$mm{^2}$ (Scale bar: 100 $\mu m/h$).}
    \end{figure}
    
    \begin{figure}
    \includegraphics[width=0.5\textwidth]{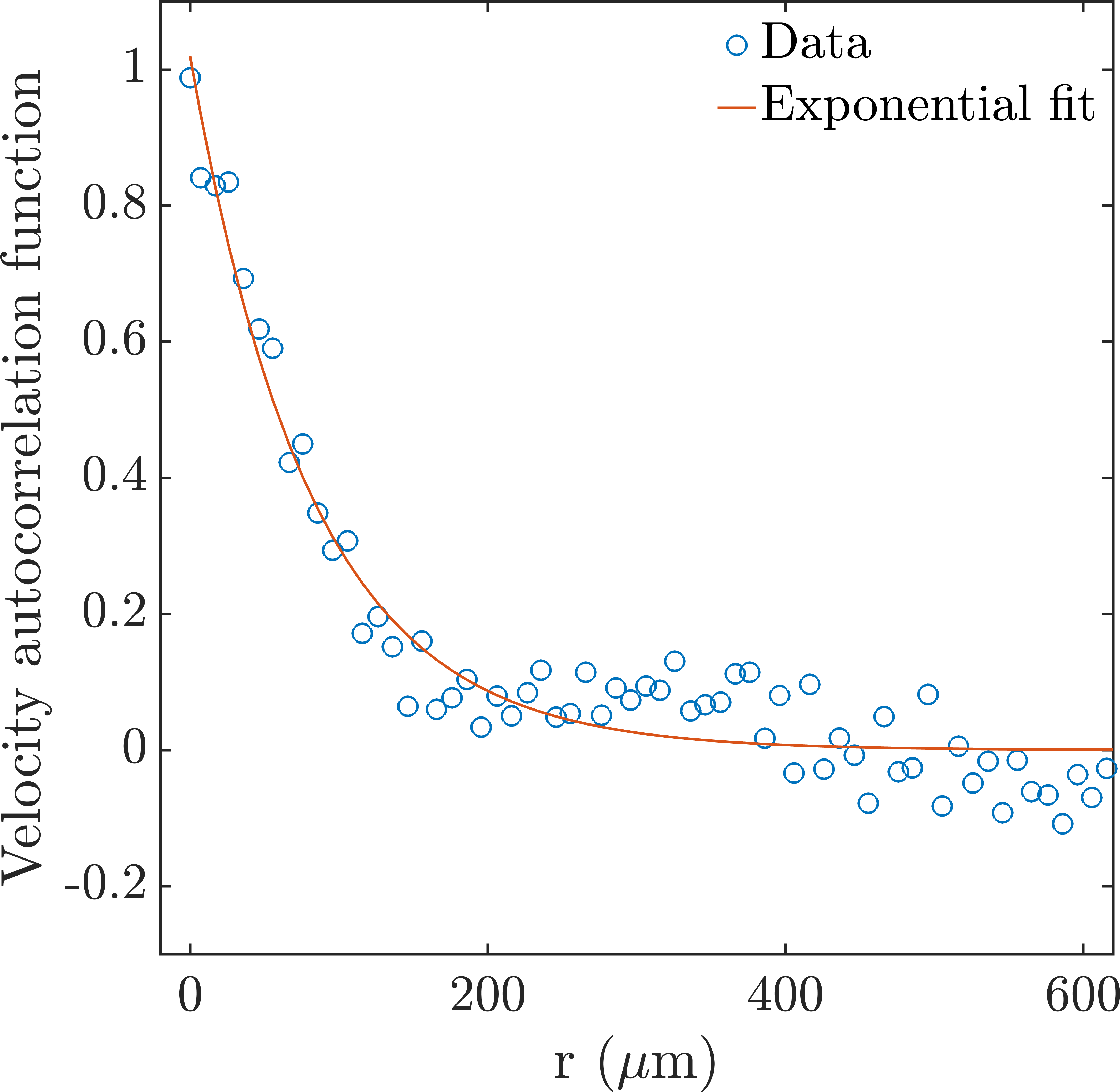}
    \caption{\label{vel_corr_fun}  
    (i) A typical graph of the spatial correlation function of tangential velocity with an exponential fit.}
    \end{figure}
    
    
    \clearpage
    \bibliography{SI_ref}